\documentclass[a4paper,fleqn,usenatbib]{mnras}

\usepackage{newtxtext,newtxmath}

\usepackage[T1]{fontenc}
\usepackage{ae,aecompl}
\usepackage{graphicx}	
\usepackage{amsmath}	
\usepackage{amssymb}	
\usepackage{pdflscape}	

\title[Mid-IR Polarization of HAeBe Discs]{Mid-Infrared Polarization of Herbig Ae/Be Discs}

\author[D. Li et al.]{Dan Li,$^{1}$\thanks{E-mail: \href{mailto:danli1@sas.upenn.edu}{danli1@sas.upenn.edu}}\thanks{Present address: Department of Physics \& Astronomy, University of Pennsylvania, 209 South 33rd Street, Philadelphia, PA19104, USA}
Charles M. Telesco,$^{1}$
Han Zhang,$^{1}$
Christopher M. Wright,$^{2}$
\newauthor
Eric Pantin,$^{3}$
Peter J. Barnes$^{1,4}$
and Chris Packham$^{5,6}$
\\
$^{1}$Department of Astronomy, University of Florida, Gainesville, FL 32611, USA\\
$^{2}$School of Physical, Environmental, and Mathematical Sciences, University of New South Wales, Canberra, ACT 2610, Australia\\
$^{3}$Service d'Astrophysique CEA Saclay, France\\
$^{4}$School of Science and Technology, University of New England, Armidale, NSW 2351, Australia\\
$^{5}$Physics and Astronomy Department, University of Texas at San Antonio, 1 UTSA Circle, San Antonio, TX 78249, USA\\
$^{6}$National Astronomical Observatory of Japan, 2-21-1 Osawa, Mitaka, Tokyo 181-8588, Japan
}

\date{Accepted 2017 August 25. 2017 August 24; in original form 2017 July 11}

\pubyear{2017}

\begin{document}
\label{firstpage}
\pagerange{\pageref{firstpage}--\pageref{lastpage}}
\maketitle

\begin{abstract}
We measured mid-infrared polarization of protoplanetary discs to gain new insight into their magnetic fields. Using CanariCam at the 10.4 m Gran Telescopio Canarias, we detected linear polarization at 8.7, 10.3, and 12.5 $\micron$ from discs around eight Herbig Ae/Be stars and one T-Tauri star. We analyzed polarimetric properties of each object to find out the most likely interpretation of the data. While the observed mid-infrared polarization from most objects is consistent with polarized emission and/or absorption arising from aligned dust particles, we cannot rule out polarization due to dust scattering for a few objects in our sample. For those objects for which polarization can be explained by polarized emission and/or absorption, we examined how the derived magnetic field structure correlates with the disc position angle and inclination. We found no preference for a certain type of magnetic field. Instead, various configurations (toroidal, poloidal, or complex) are inferred from the observations. The detection rate (64 per cent) of polarized mid-infrared emission and/or absorption supports the expectation that magnetic fields and suitable conditions for grain alignment are common in protoplanetary discs around Herbig Ae/Be stars.
\end{abstract}


\section{Introduction}
Even though the effects of magnetic fields (B-fields) in the formation of stars and circumstellar discs are still not fully understood, there has been a steadily increasing body of observational constraints on the properties of B-fields in star-forming regions and molecular clumps/cores \citep{crutcher2012,lihb2014}. However, this -- albeit slow -- progress has not been shared in the area of protoplanetary discs, the birth places of planets, where almost no direct evidence exists bearing on their B-fields. Polarized emission from aligned grains is a good proxy for mapping out B-fields, as has been demonstrated for several young stellar objects \citep[YSOs; e.g.,][]{qiu2013,zhang2014,davidson2014,segura-cox2015,rao2014,liu2016}. To date, most observations have been conducted primarily in the millimeter range, focusing on bright Class 0-I YSOs plus a small handful of protoplanetary discs (i.e., Class II YSOs) \citep{hughes2009,hughes2013,stephens2014,kataoka2016}. The translation of mm/sub-mm polarization maps into the B-field morphology (projected on the plane-of-sky) was once thought to be straightforward, but recent studies have emphasized that scattered emission can also contribute to the observed mm/sub-mm polarization if dust grains much larger than their ISM (interstellar medium) counterparts are present in discs \citep{kataoka2015,yang2016,tazaki2017}. Ambiguities in the interpretation of mm/sub-mm data arising from scattered polarization can be mitigated by joint infrared-millimeter observations \citep{tazaki2017}.

Here we present a mid-infrared (mid-IR) study of a small sample of pre-main sequence stars using CanariCam \citep{telesco2003,packham2005}, the mid-IR multi-mode facility instrument of the 10.4 m Gran Telescopio Canarias (GTC) on La Palma, Spain. All targets except one in our sample are Herbig Ae/Be (HAeBe) stars (i.e., pre-main sequence stars of 2--8 $M_{\sun}$). HAeBe stars generally exhibit modest-to-strong IR excesses originating from circumstellar discs \citep{meeus2001,alonso-albi2009}. HAeBe stars within a few hundred pc are usually bright mid-IR sources, making them suitable targets for polarimetric observations. Due to the limited sensitivity of ground-based mid-IR observations, with the exception of AB Aur, we have only detected the brightest and most compact inner regions of the discs, which are unresolved in our total intensity $I$ or polarized intensity $PI$ images. However, we show that mid-IR polarimetry provides useful constraints on the B-fields even for unresolved mid-IR-emitting discs.

In Section~\ref{sec:observations} we describe the sample and our observations. In Sections \ref{sec:results} and \ref{sec:analyses} we present and analyze the data. In Sections \ref{sec:discussion} and \ref{sec:summary} we discuss our findings and summarize the main conclusions from this work.

\defcitealias{li2016}{L16}

\section{Observations}\label{sec:observations}
\subsection{The Sample}
We selected 10 HAeBe stars and one T-Tauri star for the present study (Table~\ref{tab:sample}). All targets are bright (a few Jy or brighter) near 10 $\micron$. We only selected sources for which the disc inclinations and orientations were available from the literature (Table~\ref{tab:angles}). Our sample virtually contains most such targets readily observable (i.e., with airmass $<$ 2) from the GTC. There is considerable diversity even within this small sample, which includes both very young (less than 1 Myr) and relatively old (a few Myr) systems that are either highly or moderately inclined, or nearly face-on. HL Tau, the only T-Tauri star in our sample, is a well studied Class I object and one of the brightest T Tauri stars in the mid-IR with abundant geometric information available through ALMA observations \citep{alma2015}. Moreover, HL Tau is among the few protoplanetary discs for which mm/sub-mm polarization has been observed and modeled \citep{stephens2014,yang2016}. We also include here our observations of AB Aur, which were presented elsewhere \citep[][hereafter \citetalias{li2016}]{li2016} and which provide a useful comparison for other unresolved sources in the current sample.

\begin{table*}
\caption{Stellar properties.}
\label{tab:sample}
\begin{tabular}{lcccccccl} 
\hline
Object & Alternative & Spectral & Distance & log $L$ & $M$ & Age & $A_{V}$ & Ref. \\
& Name & Type & (pc) & ($L_{\sun}$) & ($M_{\sun}$) & (Myr) & (mag) & \\
\hline
MWC 1080A &                 & B0     & 1000  & 4.59  & 20.6 & 0.22 & 5.3        & 1,2 \\
MWC 297     &                 & B1.5  & 250    & 4.47  & 10    & 1       & 8          & 3 \\
HD 200775  & MWC 361 &  B2.5 & 430   & 3.89   & 10    & 1.6    & 1.9      & 4 \\
VV Ser         &                  & B6     & 614   & 2.5     & 3.95 & 1.2    & 3.0        & 5,6 \\
HD 179218  & MWC 614 & A0     & 200   & 1.99   & 3.04 & 1.06  & 0.4       & 7,8 \\
AB Aur         & HD 31293 & A1     & 144   & 1.84   & 2.77 & 1.74  & 0.9        & 8--10  \\
HD 163296  & MWC 275 & A1     & 122   & 1.53   & 2.49 & 3.0    & 0.22      & 7--9 \\
MWC 480    & HD 31648 & A4    & 137   & 1.34   & 1.97 & 6.7    & 0.1     & 6,9,11 \\
MWC 758    & HD 36112 & A5     & 205   & 1.45    & 2.17 & 4.17  & 0.5      &  8,9,12 \\
CQ Tau        & HD 36910 & A8     & 140   & 1.2     & 1.5   & 10     & -- & 9,13 \\
HL Tau         &                  & K5     & 140   & 0.18   & 0.7   & 1         & 24--33  & 14,15 \\
\hline
\multicolumn{9}{l}{References: (1) \citealt{hillenbrand1992}; (2) \citealt{levreault1988}; (3) \citealt{drew1997};}\\
\multicolumn{9}{l}{(4) \citealt{vandenancker1998}; (5) \citealt{pontoppidan2007_vvser2}; (6) \citealt{montesinos2009};}\\
\multicolumn{9}{l}{(7) \citealt{mora2001}; (8) \citealt{manoj2006}; (9) \citealt{perryman1997};}\\
\multicolumn{9}{l}{(10) \citealt{hernandez2004}; (11) \citealt{liu2011}; (12) \citealt{vanboekel2005}; }\\
\multicolumn{9}{l}{(13) \citealt{testi2003}; (14) \citealt{rebull2004}; (15) \citealt{white2004}.}
\end{tabular}
\end{table*}

\begin{table*}
\caption{Properties of discs and interstellar B-fields.}
\label{tab:angles}
\begin{tabular}{lccclcl} 
\hline
Object & 10-$\micron$ Silicate & Disc P.A. & Inclination & Ref. & IS B-field P.A.$^a$ & Ref.\\
 & Feature Type & (degree) & (degree) &  & (degree) & \\
\hline
MWC 1080A & Abs & 135 $\pm$ 5 & 55 $\pm$ 5 & 1,2 & 70 & 19 \\
MWC 297     & Abs & 165 $\pm$ 15 & 13 $\pm$ 5 & 1,3 & 40 & 19,20 \\
HD 200775   & Em &  6.9 $\pm$ 1.5 & 54.5 $\pm$ 1.2 & 1,4 & 142 & 19 \\
VV Ser          & Em & 15 $\pm$ 5 & 72 $\pm$ 5 & 5 & 60 & 20,21 \\
HD 179218   & Em & 22 $\pm$ 3 & 56 $\pm$ 2 & 1,9 & 4 & 20 \\
AB Aur          & Em & 70 $\pm$ 10 & 27 $\pm$ 5 & 6--8 & 70 & 19,22 \\
HD 163296   & Em & 136 $\pm$ 2 & 48 $\pm$ 2 & 10 & 175 & 23 \\
MWC 480     & Em & 148 $\pm$ 2 & 38 $\pm$ 5 & 11--13 & -- &  \\
MWC 758     & Em & 65 $\pm$ 7 & 21 $\pm$ 2 & 14 & -- &  \\
CQ Tau & Em & 54 $\pm$ 1 & 29 $\pm$ 2 & 15,16 & 42 & 24 \\
HL Tau & Abs & 138.02 $\pm$ 0.07 & 46.72 $\pm$ 0.05 & 17,18 & 77 & 25 \\
\hline
\multicolumn{7}{l}{$^a$ Interstellar B-field P.A.s inferred from optical observations on nearby field stars.}\\
\multicolumn{7}{l}{References: (1) \citealt{acke2004iso}; (2) \citealt{li2014}; (3) \citealt{manoj2007}; }\\
\multicolumn{7}{l}{(4) \citealt{okamoto2009}; (5) \citealt{pontoppidan2007_vvser2}; (6) \citealt{vanboekel2005}; }\\
\multicolumn{7}{l}{(7) \citealt{hashimoto2011}; (8) \citealt{tang2012}; (9) \citealt{fedele2008};}\\
\multicolumn{7}{l}{(10) \citealt{tannirkulam2008}; (11) \citealt{pietu2007}; (12) \citealt{grady2010}; }\\
\multicolumn{7}{l}{(13) \citealt{kusakabe2012}; (14) \citealt{isella2010}; (15) \citealt{doucet2006};}\\
\multicolumn{7}{l}{(16) \citealt{chapillon2008}; (17) \citealt{alma2015}; (18) \citealt{kospal2012};}\\
\multicolumn{7}{l}{(19) \citealt{hillenbrand1992}; (20) \citealt{rodrigues2009}; (21) \citealt{rostopchina2000};}\\
\multicolumn{7}{l}{(22) \citealt{pontefract2000}; (23) \citealt{yudin2000}; (24) \citealt{berdyugin1990}; }\\
\multicolumn{7}{l}{(25) \citealt{menard2004}.}\\
\end{tabular}
\end{table*}
 
\subsection{Observations and Data Reduction}
Polarimetric images were obtained between 2013 and 2016 using CanariCam. Three filters, \textit{Si}-2 ($\lambda=8.7$ $\micron$, $\Delta\lambda$ = 1.1 $\micron$), \textit{Si}-4 ($\lambda=10.3$ $\micron$, $\Delta\lambda$ = 0.9 $\micron$), and \textit{Si}-6 ($\lambda=12.5$ $\micron$, $\Delta\lambda$ = 0.7 $\micron$), were selected to sample the mid-IR polarization across the 10-$\micron$ silicate feature. Using the methodology described in detail by \citet[]{aitken2004} (see also \citetalias{li2016} and \citealt{barnes2015}), we can use this combination of filters to ascertain polarization contributions arising from non-spherical emitting dust particles and non-spherical absorbing particles. 

The typical integration time was 900 s (in three 300-s exposures) in each filter. The angular resolution was $\sim$0.3--0.5 arcsec, depending on the filter and seeing conditions (Table~\ref{tab:log}). We employed the standard chop-nod procedures during each integration to correct for the thermal background of the sky and the telescope. Each observation on a sample target was followed by an observation on a Cohen mid-IR standard \citep{cohen1999}, usually a bright (a few tens of Jy at 10 $\micron$) K III star within 10\degr\, of the sample target. This Cohen star served as the photometric standard, the point-spread-function (PSF) reference, as well as the instrumental polarization reference.

The observations were made using CanariCam's dual-beam polarimeter mode that utilizes a Wollaston prism to separate incoming light into the ordinary and extraordinary beams, which are recorded by the detector array simultaneously. We reduced the data using \sc{idealcam} \rm \citep{li2013idealcam}, which computed the Stokes \textit{I}, \textit{Q}, and \textit{U} images (and uncertainties) from raw CanariCam data. Instrumental polarization, which was measured to be around 0.5--0.8 per cent depending on the filter and which is well characterized and reproducible for CanariCam, was subtracted from the observations in the $Q$--$U$ plane.

\section{Results}\label{sec:results}
Among our targets only AB Aur is resolved in both $I$ and $PI$ images, as discussed in detail in \citetalias{li2016}. The protoplanetary disc of AB Aur is clearly extended in the \textit{I}, \textit{Q}, and \textit{U} images. For other unresolved targets, linear polarization was measured with circular apertures centred on the stars. The radius of the aperture was set equal to the angular resolution (i.e., the FWHM of the PSF as listed in Table~\ref{tab:log}). The linear polarization $p$ is calculated as $\sqrt{Q^2+U^2-\sigma^2}/I$, where $\sigma$ is the noise (standard deviation) of $Q$ and $U$ measured from un-stacked frames in the raw CanariCam data, and serves to remove the positive bias in $p$ resulting from the noise. The polarization position angle (P.A., measured east of north) is calculated as $0.5\,\mathrm{arctan}(U,\,Q)$, where the $\mathrm{arctan}(U,\,Q)$ function is used to compute $\mathrm{arctan}(U/Q)$ avoiding the $\pi$ ambiguity. Including AB Aur, we detected polarization ($\geqslant$ 3$\sigma$) in nine of the eleven objects (Table~\ref{tab:results}). The polarization at 10.3 $\micron$ ranges from 0.22 per cent (MWC 297) to 2.74 per cent (CQ Tau). The unweighted mean polarization of the sample (excluding the non-detections of MWC 480 and MWC 758) is 0.5, 0.86, and 0.88 per cent at 8.7, 10.3, and 12.5 $\micron$, respectively. 

\begin{table*}
\caption{Observations.}
\label{tab:log}
\begin{tabular}{lcccccl} 
\hline
Object & $\lambda$ & UT Date & Integration$^a$ & Airmass & FWHM$^b$ & Flux/PSF  \\
 & ($\micron$) & (yyyy-mm-dd) & (s) & & (arcsec) & Calibrator \\
\hline
VV Ser & 8.7 & 2013-08-04 & 300 $\times$ 3 & 1.18 & 0.24 & $\eta$ Ser  \\
& 10.3 & 2013-08-04 & 300 $\times$ 3 & 1.14 & 0.27 &  \\
& 12.5 & 2013-08-05 & 300 $\times$ 3 & 1.58 & 0.33 &    \\

MWC 1080A & 8.7 & 2013-08-05 & 300 $\times$ 3 & 1.20 & 0.29 & HD 1240  \\
& 10.3 & 2013-08-05 & 300 $\times$ 3 & 1.18 & 0.26  &  \\
& 12.5 & 2013-08-06 & 300 $\times$ 3 & 1.26 & 0.31  &  \\
							      
HD 179218  & 8.7 & 2013-08-07 & 300 $\times$ 3 & 1.39 & 0.37 & $\alpha$ Vul  \\
& 10.3 & 2013-08-07 & 300 $\times$ 3 & 1.20 & 0.36  & \\
& 12.5 & 2013-08-05 & 300 $\times$ 3 & 1.15 & 0.35  &  \\
							      
HD 200775  & 8.7 & 2013-09-10 & 300 $\times$ 3 & 1.48 & 0.32 & HD 198149  \\
& 10.3 & 2013-09-10 & 300 $\times$ 3 & 1.37 & 0.26  &   \\
& 12.5 & 2013-09-10 & 300 $\times$ 3 & 1.32 & 0.38  &   \\

MWC 480  & 12.5 & 2014-01-11 & 300 $\times$ 2 & 1.05 & 0.33 &  HD 31398  \\

MWC 758 & 8.7 & 2014-01-12 & 300 $\times$ 4 & 1.10 & 0.30 & HD 31398  \\
& 10.3 & 2014-01-12 & 300 $\times$ 4 & 1.26 & 0.34  & \\
& 12.5 & 2014-01-11 & 300 $\times$ 3 & 1.29 & 0.38  &  \\							   
                                                                
MWC 297  & 8.7 & 2014-05-27 & 300 $\times$ 2 & 1.19 & 0.47  & HD 168723  \\
& 10.3 & 2014-05-27 & 300 $\times$ 2 & 1.23 & 0.41  &   \\
& 12.5 & 2014-06-06 & 300 $\times$ 2 & 1.45 & 0.41  &   \\

AB Aur & 8.7 &  2015-03-06 & 360 $\times$ 2 & 1.12 & 0.36 &  HD 31398  \\
& 10.3 & 2015-02-06 & 360 $\times$ 3 & 1.28 & 0.35 &  \\
& 12.5 & 2015-02-05 & 320 $\times$ 3 & 1.19 & 0.34 &  \\

CQ Tau
& 8.7 &  2015-03-30 & 300 $\times$ 2 & 1.48 & 0.43 &  HD 34334  \\
& 10.3 & 2015-03-30 & 300 $\times$ 3 & 1.19 & 0.41 &  \\
& 12.5 & 2016-01-22 & 320 $\times$ 3 & 1.56 & 0.50 &  \\

HD 163296
& 8.7 & 2015-07-01 & 300 & 2.02 & 0.38 & HD 169916 \\
& 10.3 & 2015-07-01 & 300 $\times$ 2 & 1.82 & 0.43 & \\
& 12.5 & 2015-06-08 & 300 & 1.60 & 0.41 & \\

HL Tau
& 8.7 &  2016-02-17 & 300 $\times$ 2 & 1.08 & 0.30 &  HD 28305  \\
& 10.3 & 2016-01-25 & 365 $\times$ 3 & 1.08 & 0.32 &  \\
& 12.5 & 2016-03-17 & 320 $\times$ 3 & 1.41 & 0.32 &  	\\
\hline
\multicolumn{7}{l}{$^a$ On-source integration time.}\\
\multicolumn{7}{l}{$^b$ FWHM of the PSF calibrator.}\\
\end{tabular}
\end{table*}

\begin{table}
\caption{Aperture photometry and polarimetry.}
\label{tab:results}
\begin{tabular}{lccccc} 
\hline
Object & $\lambda$ & $F_{\nu}$ & $p$ & $\theta$ \\
 & ($\micron$) & (Jy) & (per cent) & (degree) \\ 
\hline
MWC 1080A 
& 8.7  & 15.7 $\pm$ 1.6 & 0.12 $\pm$ 0.03 & 110 $\pm$ 1 \\
& 10.3 & 13.4 $\pm$ 1.3 & 0.29 $\pm$ 0.03 & 104 $\pm$ 1 \\
& 12.5  & 19.2 $\pm$ 1.9 & 0.27 $\pm$ 0.07 & 71$\pm$ 10 \\

MWC 297 
& 8.7  & 115.9 $\pm$ 11.6 & 0.08 $\pm$ 0.05 & 80 $\pm$ 19 \\
& 10.3  & 103.2 $\pm$ 10.3 & 0.22 $\pm$ 0.05 & 74 $\pm$ 7 \\
& 12.5  & 111.6 $\pm$11.2 & 0.02 $\pm$ 0.03 & 91 $\pm$ 26 \\

HD 200775 
& 8.7  & 7.8 $\pm$ 0.8 & 0.47 $\pm$ 0.09 & 105 $\pm$ 2 \\
& 10.3  & 7.3 $\pm$ 0.7 & 0.52 $\pm$ 0.15 & 110 $\pm$ 1 \\
& 12.5  & 5.3 $\pm$ 0.5 & 1.2 $\pm$ 0.1 & 111 $\pm$ 1 \\

VV Ser 
& 8.7  & 3.6 $\pm$ 0.4 & 0.48 $\pm$ 0.04 & 105 $\pm$ 1  \\
& 10.3 & 4.2 $\pm$ 0.4 & 0.54 $\pm$ 0.04 & 97 $\pm$ 2  \\
& 12.5 & 4.0 $\pm$ 0.4 & 1.0 $\pm$ 0.3 & 96 $\pm$ 6 \\

HD 179218 
& 8.7  & 16.1 $\pm$ 1.6 & 0.41 $\pm$ 0.03 & 113 $\pm$ 1 \\
& 10.3 & 23.7 $\pm$ 2.4 & 0.78 $\pm$ 0.02 & 111 $\pm$ 1 \\
& 12.5 & 16.8 $\pm$ 1.7 & 0.81 $\pm$ 0.05 & 112 $\pm$ 1 \\

AB Aur
& 8.7 & 16.5 $\pm$ 1.7 & 0.25 $\pm$ 0.02 & 163 $\pm$ 4 \\
& 10.3 &  38.8 $\pm$ 3.9 & 0.50 $\pm$ 0.02 & 168 $\pm$ 1 \\
& 12.5 &  12.4 $\pm$1.2 & 0.57 $\pm$ 0.05 & 176 $\pm$ 3 \\

HD 163296
& 8.7  & 4.2 $\pm$ 0.4 & 0.63 $\pm$ 0.04 & 46 $\pm$ 2 \\
& 10.3 & 6.2 $\pm$ 0.6 & 0.84 $\pm$ 0.03 & 37 $\pm$ 1 \\
& 12.5 & 12.4 $\pm$ 1.2 & 0.58 $\pm$ 0.12 & 58 $\pm$ 9 \\

MWC 480 
& 12.5 & 6.5 $\pm$ 0.7 & $<$ 2.1$^a$ & -- \\

MWC 758 
& 8.7  & 3.7 $\pm$ 0.4 &  $<$ 2.7$^a$ & --  \\
& 10.3  & 4.6 $\pm$ 0.5 & $<$ 2.1$^a$ & --  \\
& 12.5  & 2.5 $\pm$ 0.3 & $<$ 3.0$^a$ & --  \\

CQ Tau
& 8.7 & 2.8 $\pm$ 0.3 & 1.41 $\pm$ 0.15 & 18 $\pm$ 1 \\
& 10.3 & 8.3 $\pm$ 0.8 & 2.74 $\pm$ 0.09 & 13 $\pm$ 1 \\
& 12.5 & 3.1 $\pm$ 0.3 & 2.88 $\pm$ 0.40 & 7 $\pm$ 2 \\

HL Tau
& 8.7 & 5.5 $\pm$ 0.5 & 0.65 $\pm$ 0.04 & 88 $\pm$ 2 \\
& 10.3 & 5.8 $\pm$ 0.6 & 1.30 $\pm$ 0.03 & 89 $\pm$ 1 \\
& 12.5 & 7.3 $\pm$ 0.7 & 0.62 $\pm$ 0.08 & 94 $\pm$ 3 \\
\hline
\multicolumn{6}{l}{$^a$ 3$\sigma$ upper limits.}\\
\multicolumn{6}{l}{The aperture radius for measuring polarization is 4 $\times$ FWHM}\\
\multicolumn{6}{l}{(for AB Aur) or 1 $\times$ FWHM (for all other sources), where}\\
\multicolumn{6}{l}{FWHM is measured with the PSF calibrator as shown in}\\
\multicolumn{6}{l}{Table~\ref{tab:log}. The radius for aperture photometry is 3 $\times$ FWHM}\\
\multicolumn{6}{l}{ for all sources. Photometric errors are dominated by flux}\\
\multicolumn{6}{l}{calibration uncertainties ($\pm$ 10\%). The polarization position}\\
\multicolumn{6}{l}{angle ($\theta$) is measured east of north.}\\
\end{tabular}
\end{table}

As the only resolved target in the sample, AB Aur serves as a benchmark allowing us to more effectively test our models for computing 10-$\micron$ polarization from a protoplanetary disc \citepalias{li2016}. Briefly, our key findings about AB Aur are as follows. The polarization of AB Aur at 10.3 $\micron$ is $\sim$0.44 per cent near the disc centre ($r<70$ au), rising to $\sim$1.4 per cent at larger radii ($70<r<170$ au). The polarization vectors are organized into two distinct patterns for these two regions of interest. Polarization vectors within 70 au are oriented almost uniformly at the P.A. of $\sim$165\degr, while those outside 70 au form a highly centrosymmetric, ring-like structure (Fig. 1 in \citetalias{li2016}). We modeled this observation using a Monte Carlo radiative transfer code that is able to compute polarization arising from dust scattering and thermal emission/absorption of aligned grains \citep{zhang2017aas}. We found that neither scattering nor thermal emission/absorption alone is able to explain the data. In the best-fitting model, scattered mid-IR emission dominates the polarized light in the outer disc but becomes negligible at smaller radii ($r<70$ au), where polarized thermal emission from aligned grains dominates. Our model also suggests that the B-field lines threading the disc need to be tilted from the spin axis of the disc in order to be able to reproduce the observations.

\section{Interpretation}\label{sec:analyses}
\subsection{Polarized emission and absorption}
If elongated grains are aligned in such a way that their minor (spin) axes are parallel to the B-field, as proposed by some widely considered dust alignment mechanisms such as radiative alignment torque (RAT; \citealt{lazarianhoang2007}), then the polarization vector of dust emission (absorption) would be perpendicular (parallel) to the B-field. If both warm (i.e., emitting) and cold (i.e., absorbing) grains are present along the line of sight, both processes will contribute to the observed polarization. In this case, the angle between the measured polarization and the $B_{\rm pos}$, the component of the B-field in the plane of the sky, can be any value between 0\degr\, and 90\degr. 

In general, at mid-IR wavelengths, both emissive and absorptive components can arise from complex radiative transfer in a cloud or multiple clouds along the line of sight. However, as illustrated by \citet{aitken2004}, these two components can be characterized individually to a reasonable degree under the assumption that no scattered light polarization is present. The methodology for separating emissive and absorptive components relies on the fact that the two components have different polarization spectra across the 10-$\micron$ silicate feature. Hence, to first order, one can fit the observed polarization (measured at three or more wavelengths, or with spectropolarimetry) to a combination of the two components \citep{aitken2004}. It is worth noting that this approach requires that the spectral properties of the dust in the emitting and absorbing regions are the same (or at least very similar). It also assumes that the absorptive polarization profile found for the BN Object in the Orion Molecular Cloud applies to other objects (see \citealt{aitken2004} for more discussion). Despite these limitations, the effectiveness of this method to interpret the mid-IR polarization from YSOs has been demonstrated in a number of studies to date (e.g., \citealt{smith2000}, \citealt{barnes2015}, \citetalias{li2016}, and \citealt{zhang2017}).

We applied this method to our sample, with results presented in Table~\ref{tab:separation}, in which the columns 2--6 show the fitted components when both emissive and absorptive polarization are included in the fit. The next three columns give the fit parameters for pure absorptive polarization and the last three columns for pure emissive polarization. Among the fit parameters, $p_{\rm a}$ and $p_{\rm e}$ refer to the peak polarization of the absorptive component and emissive component, respectively. For the absorptive component, this peak value occurs at 10.3 $\micron$, or 11.5 $\micron$ for the emissive component. The P.A. of each component is given by $\theta_{\rm a}$ and $\theta_{\rm e}$. The value of $\chi^2$ is used to quantify the goodness of fit.

\begin{landscape}
\begin{table}
\caption{Fits to polarimetry.}
\label{tab:separation}
\begin{tabular}{lccccccccccc} 
\hline
 & \multicolumn{5}{c}{Two-component Fit} & \multicolumn{3}{c}{Absorption Only} & \multicolumn{3}{c}{Emission Only}  \\ 
Object & $p_{\rm a}$ & $\theta_{\rm a}$ & $p_{\rm e}$ & $\theta_{\rm e}$ & $\chi^2$ & $p_{\rm a}$ & $\theta_{\rm a}$ & $\chi^2$ & $p_{\rm e}$ & $\theta_{\rm e}$ & $\chi^2$ \\
& (\%) & (\degr) & (\%) & (\degr) & & (\%) & (\degr) & & (\%) & (\degr) & \\
\hline
MWC 1080A & 0.38 $\pm$ 0.11 & --59 $\pm$ 9 & 0.20 $\pm$ 0.11 & 57 $\pm$ 15 & 11.0
& \textbf{0.29 $\pm$ 0.02} & \textbf{--75 $\pm$ 2} & \textbf{14.2} & 0.27 $\pm$ 0.02 & --77 $\pm$ 2 & 20.8 \\
MWC 297 & \textbf{0.38 $\pm$ 0.08} & \textbf{73 $\pm$ 6} & \textbf{0.17 $\pm$ 0.06} & \textbf{--20 $\pm$ 10} & \textbf{0.22}
& 0.15 $\pm$ 0.01 & 77 $\pm$ 3 & 8.9 & 0.07 $\pm$ 0.01 & 79 $\pm$ 6 & 20.5 \\
HD 200775 & \textbf{1.15 $\pm$ 0.18} & \textbf{--2 $\pm$ 4} & \textbf{1.72 $\pm$ 0.18} & \textbf{--86 $\pm$ 3} & \textbf{25.8}
& 0.66 $\pm$ 0.02 & --75 $\pm$ 1 & 121.7 & 0.69 $\pm$ 0.02 & --76 $\pm$ 1 & 66.5 \\
VV Ser & \textbf{0.97 $\pm$ 0.2} & \textbf{26 $\pm$ 7} & \textbf{1.6 $\pm$ 0.2} & \textbf{--71 $\pm$ 4} & \textbf{2.1}
& 0.65 $\pm$ 0.03 & --81 $\pm$ 1 & 44.7 & 0.68 $\pm$ 0.03 & --80 $\pm$ 1 & 18.0 \\
HD 179218 & 0.08 $\pm$ 0.10 & 62 $\pm$ 34 & 0.88 $\pm$ 0.10 & --65 $\pm$ 3 & 4.5 
& 0.84 $\pm$ 0.02 & --69 $\pm$ 1 & 81.4 & \textbf{0.86 $\pm$ 0.02} & \textbf{-68 $\pm$ 1} & \textbf{5.2} \\
AB Aur & 0.16 $\pm$ 0.08 & --56 $\pm$ 15 & 0.57 $\pm$ 0.08 & --3 $\pm$ 4 & 14.3 
& 0.53 $\pm$ 0.01 & --12 $\pm$ 1 & 60.8 & \textbf{0.55 $\pm$ 0.01} & \textbf{--12 $\pm$ 1} & \textbf{18.3} \\
HD 163296 &  \textbf{0.93 $\pm$ 0.21} &  \textbf{--7 $\pm$ 5} &  \textbf{1.36 $\pm$ 0.19} &  \textbf{61 $\pm$ 4} &  \textbf{15.5} 
& 0.94 $\pm$ 0.02 & 38 $\pm$ 1 & 72.0 & 0.96 $\pm$ 0.02 & 40 $\pm$ 1 & 36.5 \\
CQ Tau & 0.36 $\pm$ 0.58 & 18 $\pm$ 43 & 2.66 $\pm$ 0.63 & 12 $\pm$ 6 & 7.8 
& 2.87 $\pm$ 0.07 & 13 $\pm$ 1 & 26.0 & \textbf{3.02 $\pm$ 0.07} & \textbf{13 $\pm$ 1} & \textbf{8.2} \\
HL Tau & 1.03 $\pm$ 0.13 & 87 $\pm$ 4 & 0.35 $\pm$ 0.13 & --84 $\pm$ 12 & 8.2
& \textbf{1.36 $\pm$ 0.02} & \textbf{89 $\pm$ 1} & \textbf{15.0} & 1.34 $\pm$ 0.02 & 89 $\pm$ 1 & 68.1  \\
\hline
\multicolumn{12}{l}{Preferred fits are in bold font. $\chi^2$ is the sum of $\chi^2_u$ and $\chi^2_q$.}\\
\end{tabular}
\end{table}
\end{landscape}

As pointed out by \citet{aitken2004}, it is necessary to check if emissive or absorptive polarization alone can reproduce the observation reasonably well. The value of $\chi^2$ is not a conclusive criterion, because the two-component fit always yields lower $\chi^2$. For the present study, the two-component fit is favored only if it yields a $\chi^2$ value at least two times lower than that of a single-component fit. The preferred fit for each object is plotted in Fig.~\ref{fig:separation}.

\begin{figure}
\includegraphics[width=\columnwidth]{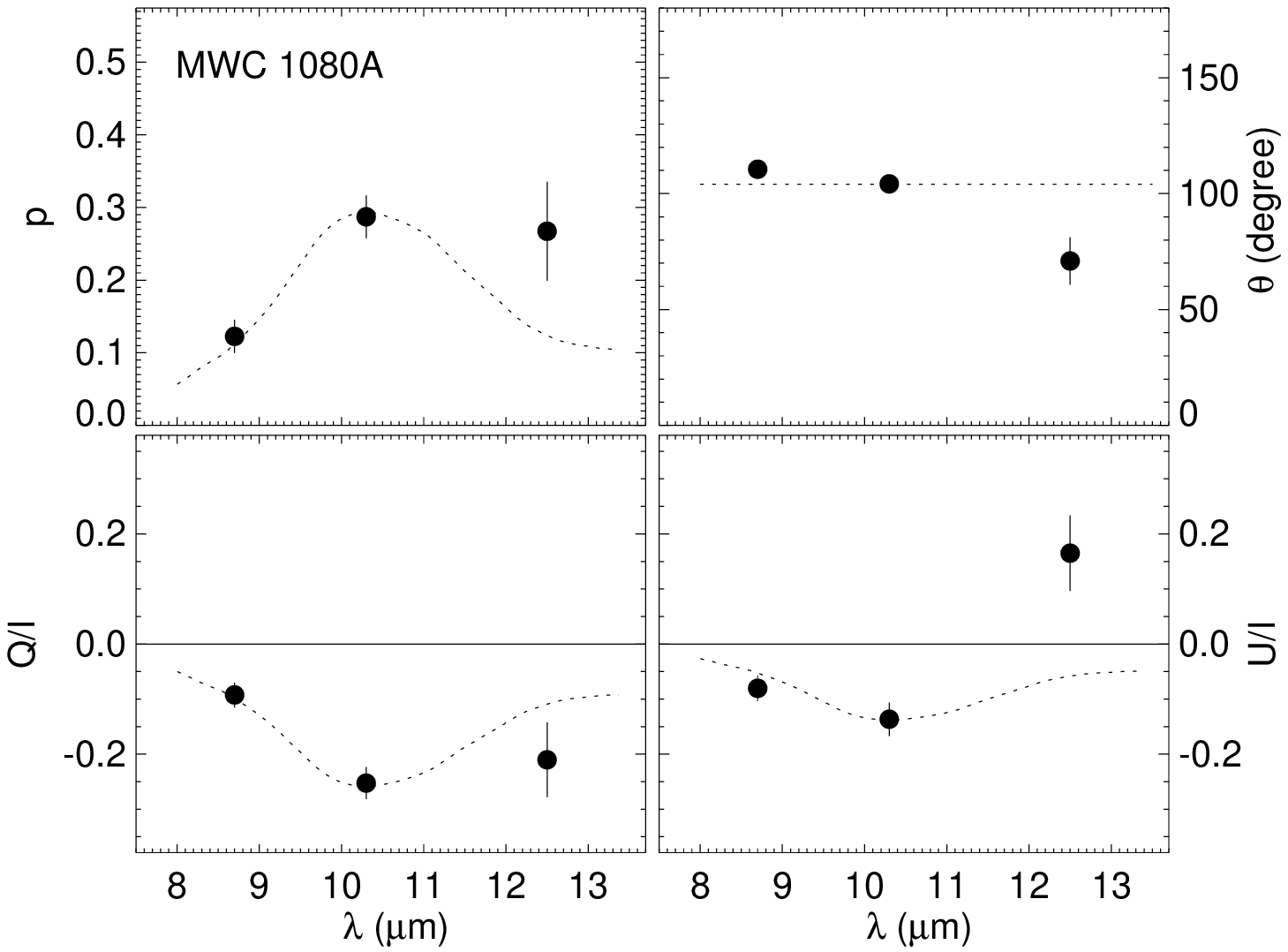}
\includegraphics[width=\columnwidth]{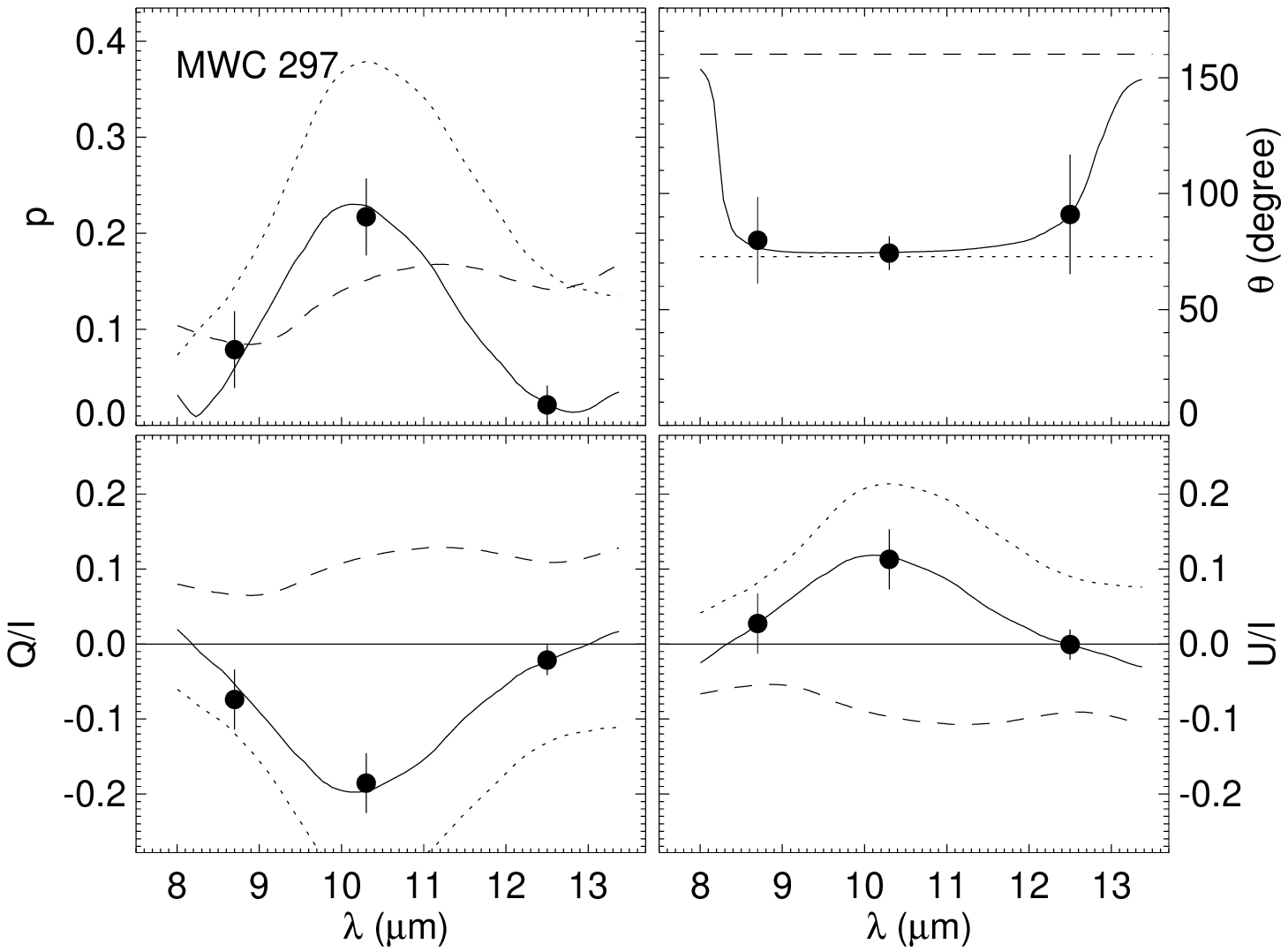}
\includegraphics[width=\columnwidth]{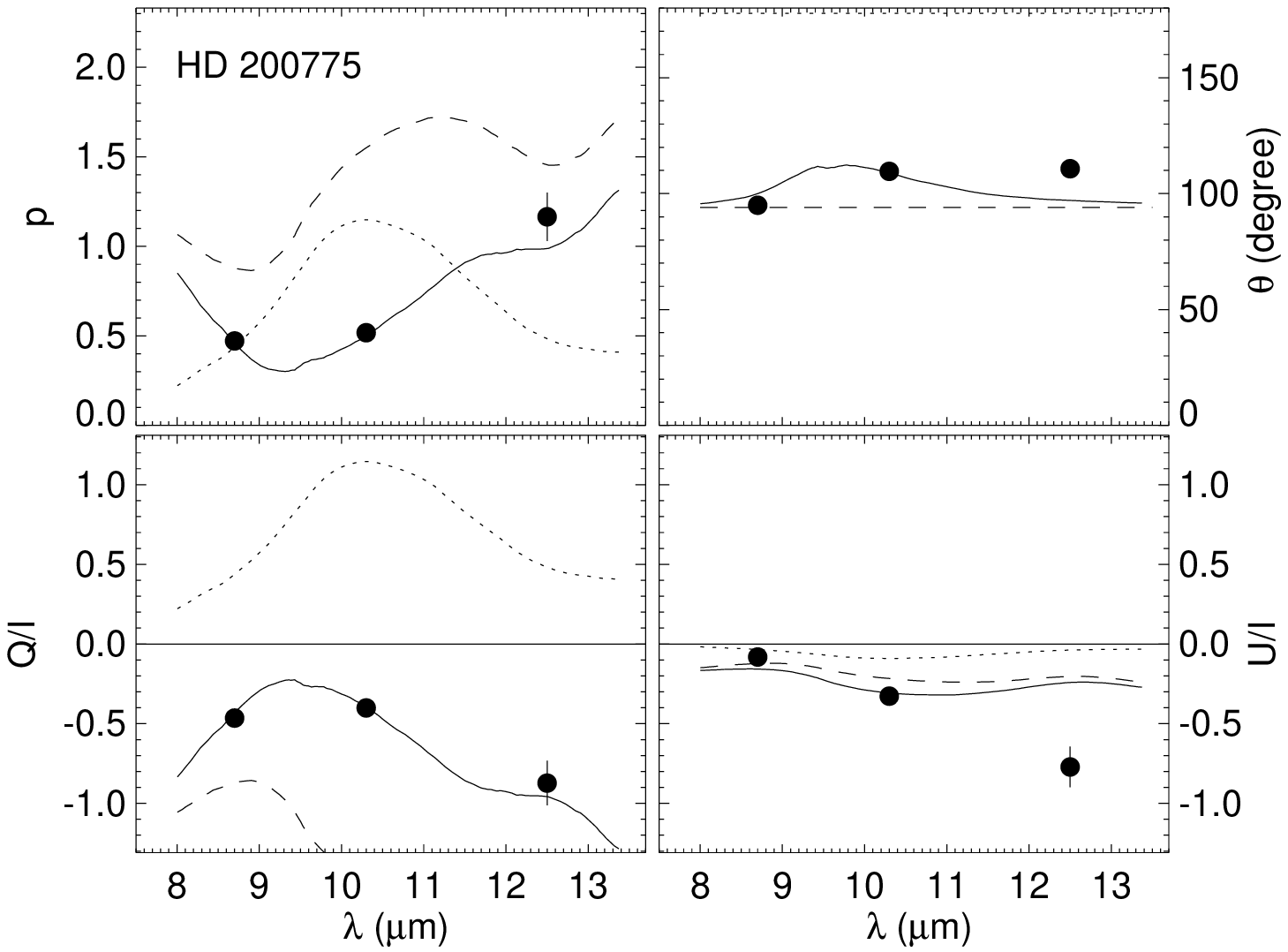}
\caption{One- or two-component polarimetric fits to measurements at 8.7, 10.3, and 12.5 $\micron$. For each object, fitting results are given in the polarization fraction ($p$, top-left), position angle ($\theta$, top-right), $Q/I$ (bottom-left), and $U/I$ (bottom-right). The dotted lines are the absorptive components, dashed lines the emissive components, and solid lines the two components combined.}
\label{fig:separation}
\end{figure}

\begin{figure}
\includegraphics[width=\columnwidth]{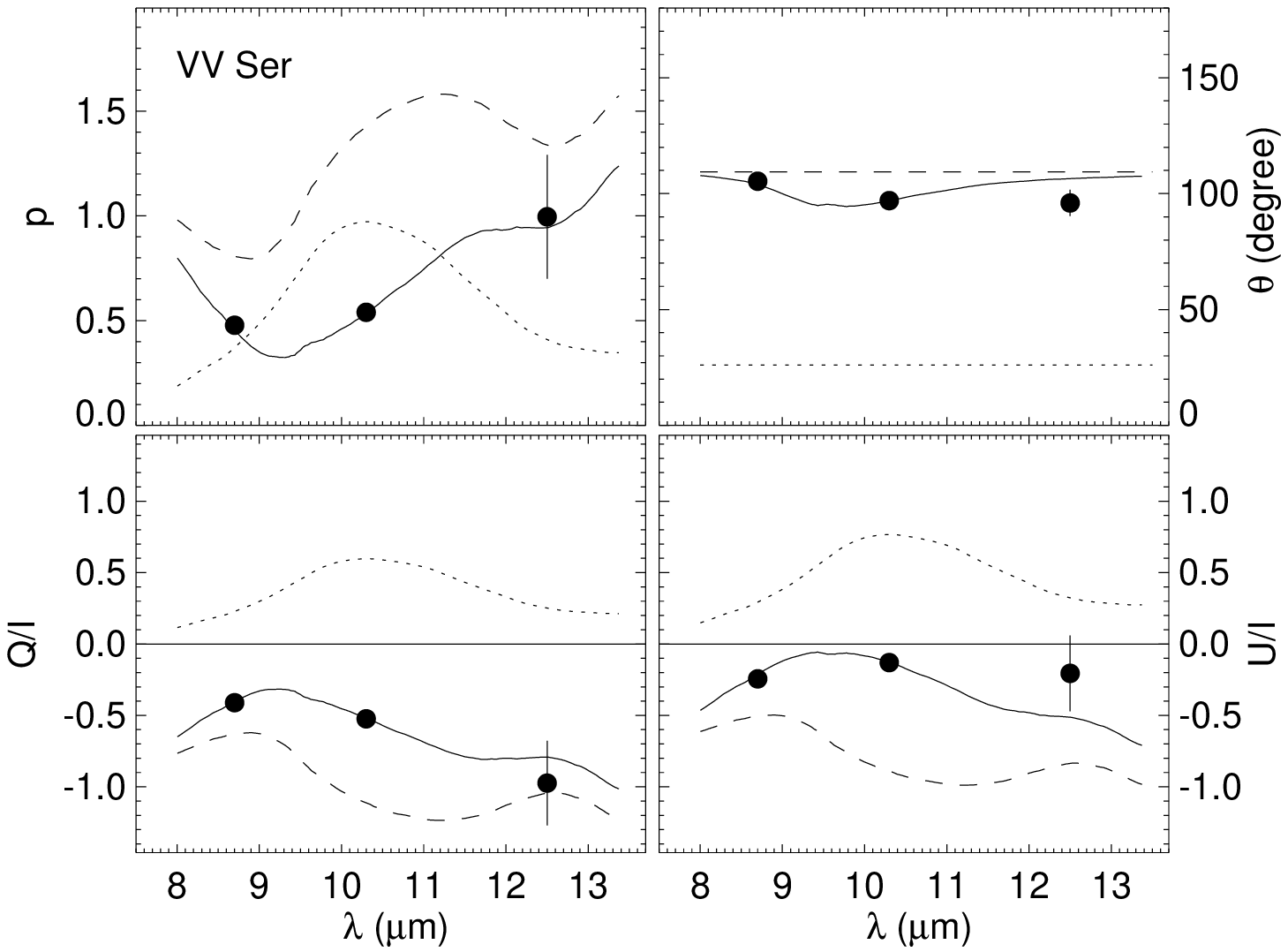}
\includegraphics[width=\columnwidth]{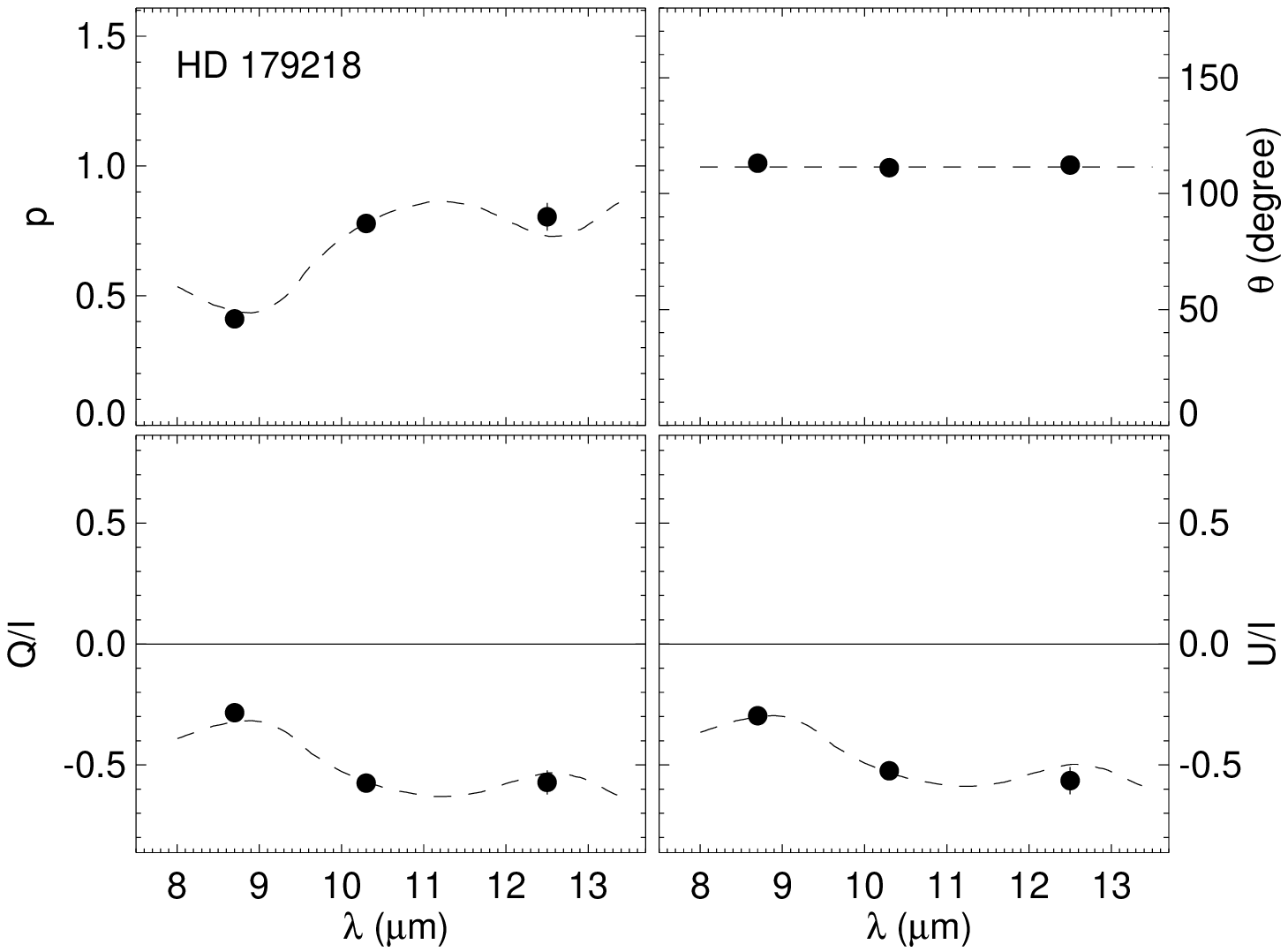}
\includegraphics[width=\columnwidth]{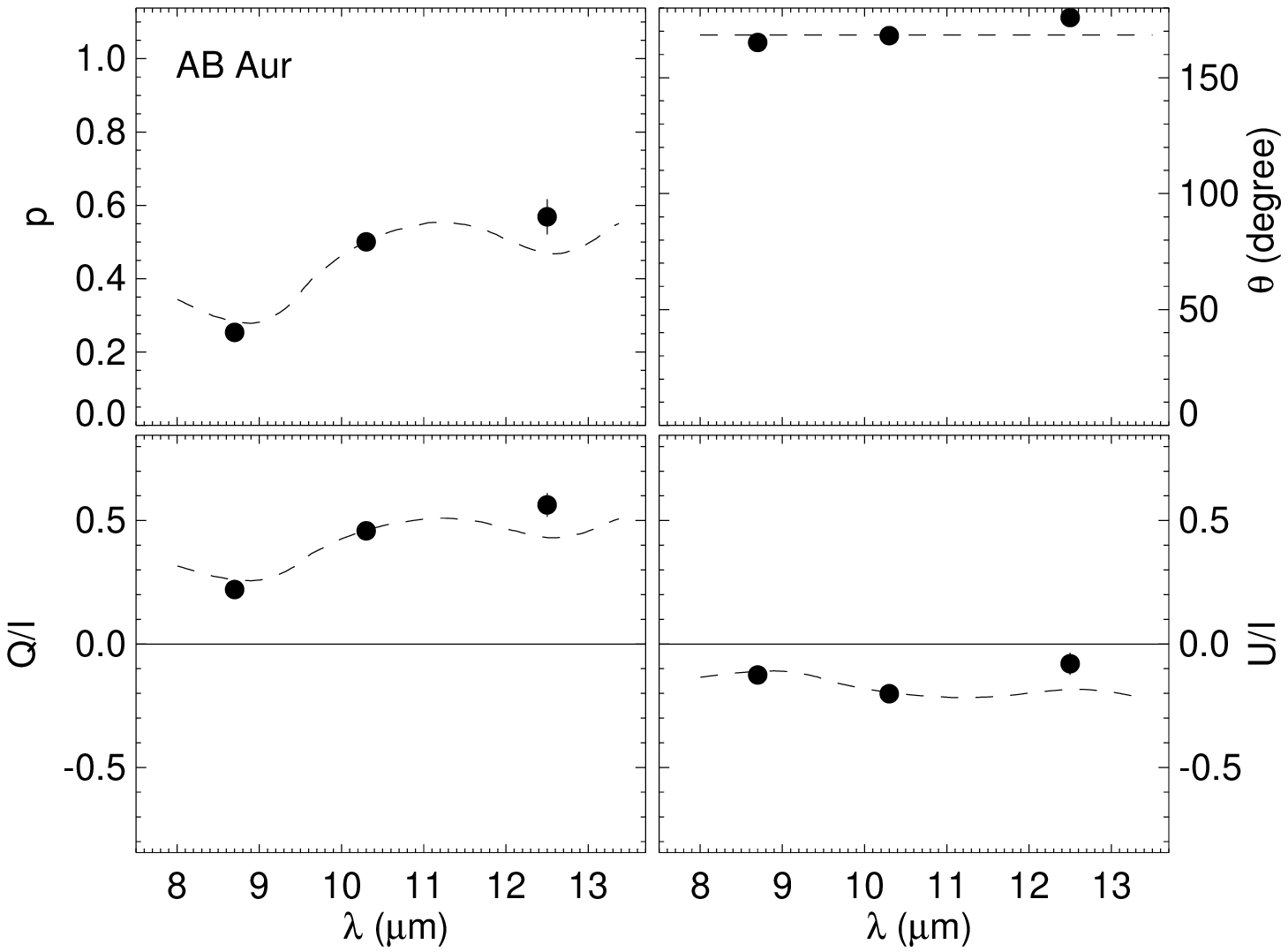}
\contcaption{One- or two-component polarimetric fits to measurements at 8.7, 10.3, and 12.5 $\micron$.}
\end{figure}

\begin{figure}
\includegraphics[width=\columnwidth]{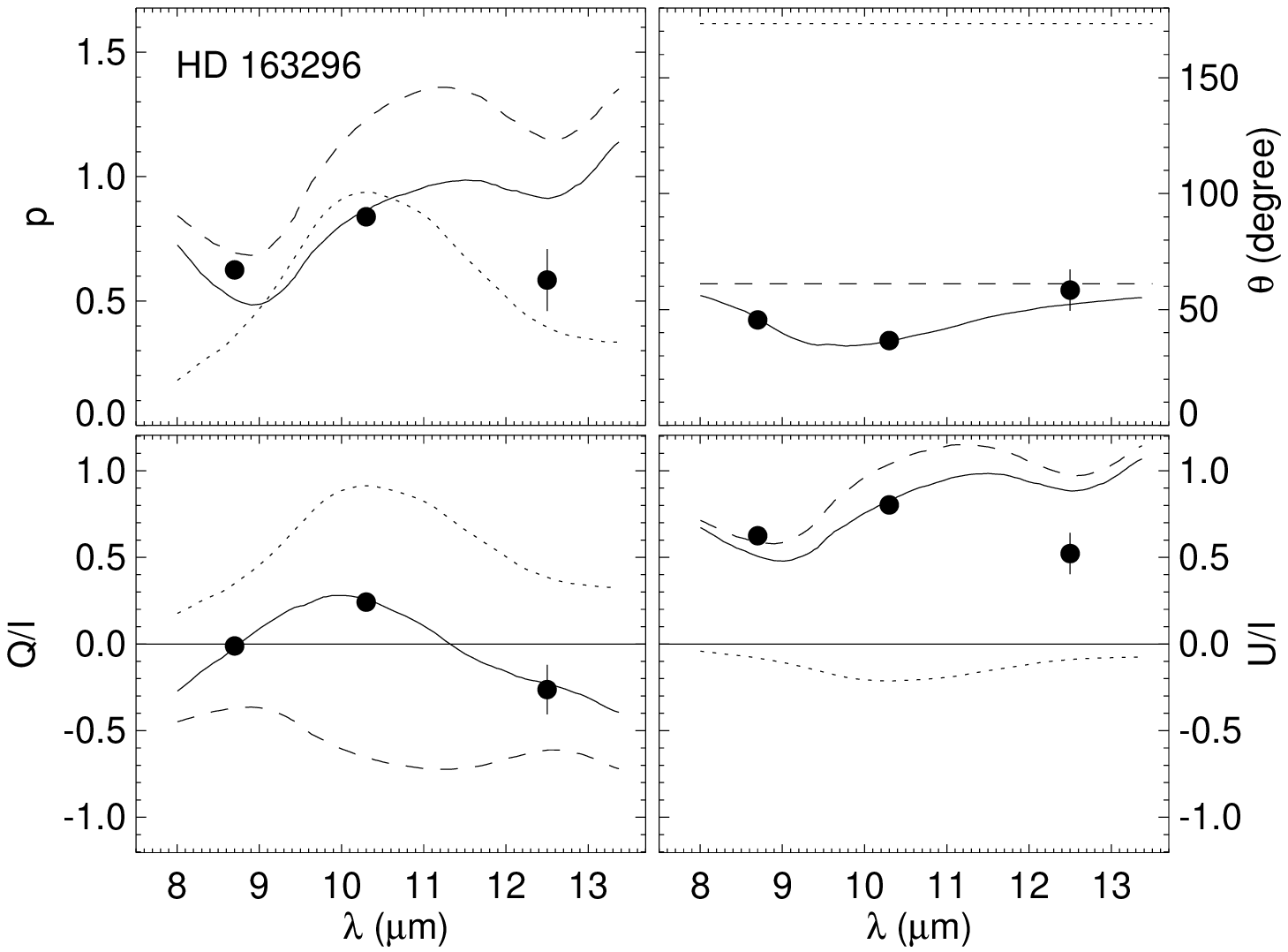}
\includegraphics[width=\columnwidth]{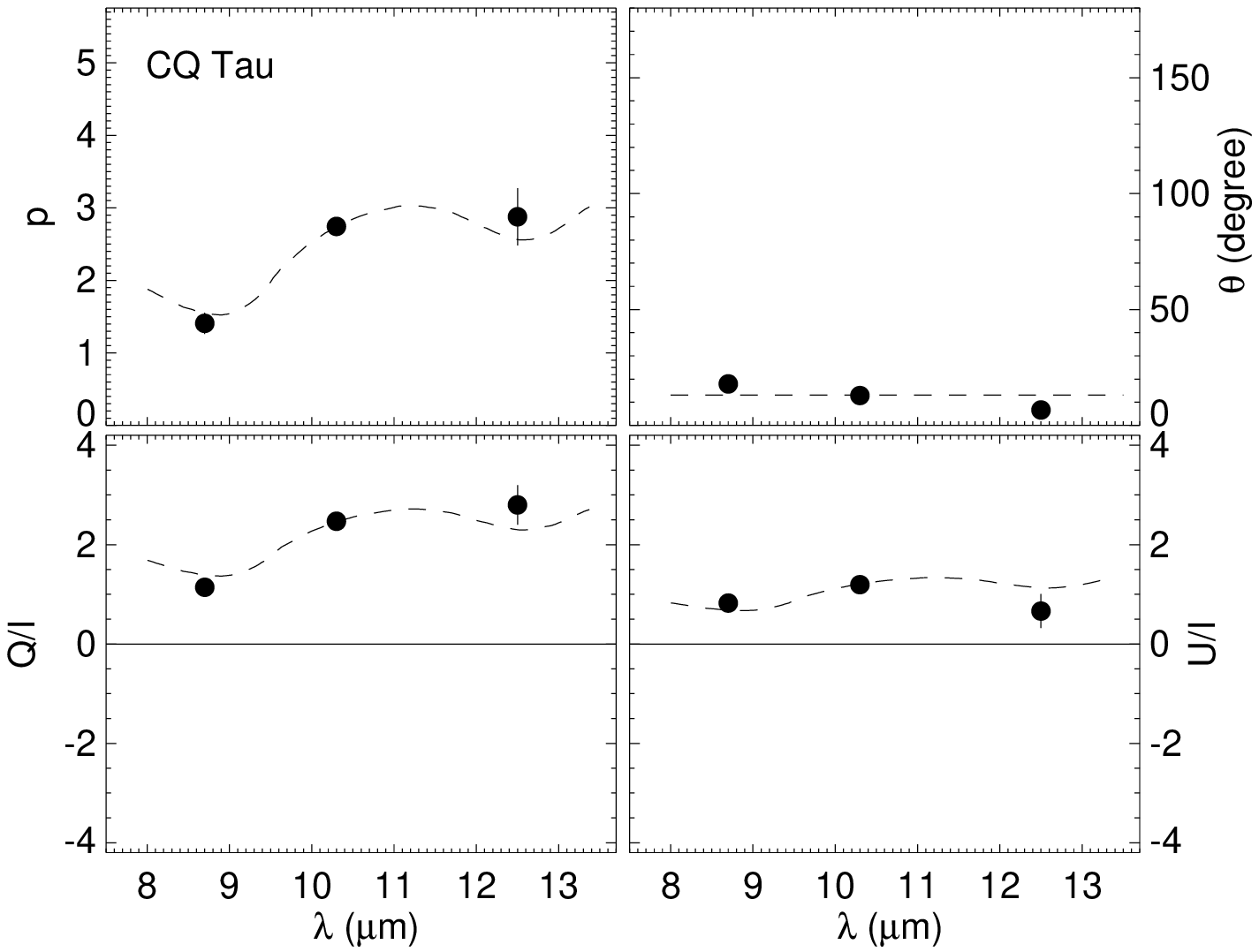}
\includegraphics[width=\columnwidth]{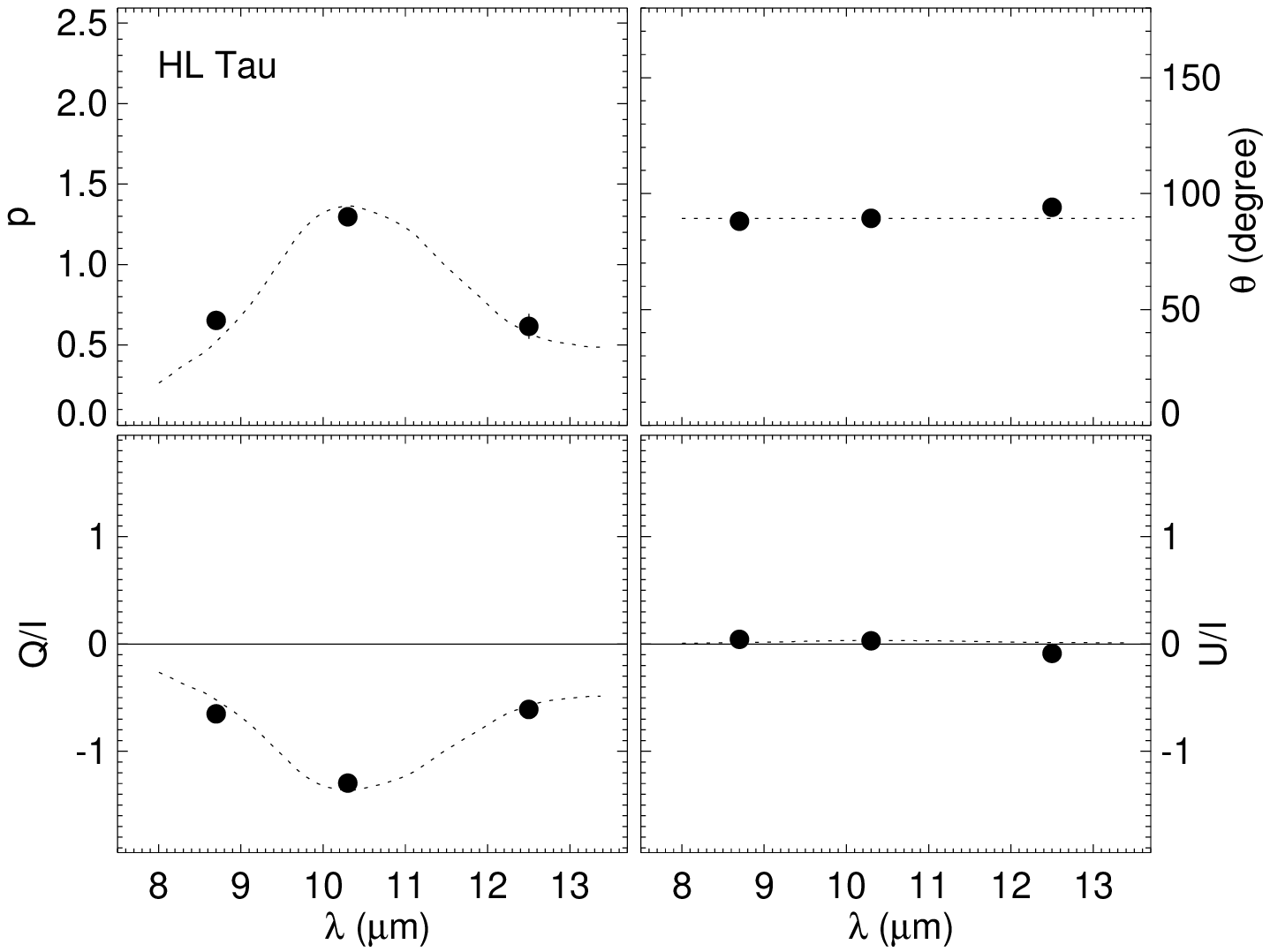}
\contcaption{One- or two-component polarimetric fits to measurements at 8.7, 10.3, and 12.5 $\micron$.}
\end{figure}

A strong correlation is observed between results of the polarimetric fits and 10-$\micron$ silicate feature shapes (Table~\ref{tab:angles}): For objects with emissive (absorptive) silicate features at 10-$\micron$, their mid-IR polarization is also mostly emissive (absorptive). It is worth noting that all four objects for which two-component fits are preferred exhibit emissive silicate features at 10 $\micron$, but the polarimetric fits suggest that their polarization is partially due to absorption, despite no hint of that component in the spectra of the total intensity. There is another apparent correlation between the visual extinction ($A_V$) and polarization: all objects with $A_V$ of a few mag or higher exhibit significant absorptive polarization.

We find that emission is the major source of mid-IR polarization for six of nine objects. Emissive polarization likely arises at the warm disc surface from which most mid-IR light is thought to be emitted \citep{chiang1997,dullemond2001}. Among the three objects for which significant absorptive polarization is inferred, MWC 1080A and MWC 297 are the two most massive stars in our sample with evidence showing that they are still associated with remnant natal clouds \citep{wang2008,alonso-albi2009,li2014}. The third object, HL Tau, is also known to be embedded in a dusty torus with $A_V>20$ mag \citep{menshchikov1999}. Therefore, it seems reasonable to attribute the absorptive polarization from these three objects to envelopes and/or toruses around the discs.

\subsection{Polarization due to scattering}
Polarimetric fits shown above assume no polarization from scattering. However, as demonstrated in \citetalias{li2016}, the contribution from scattering to the total polarization is not necessarily negligible, even though the scattered light can be orders of magnitude weaker than thermal emission from disc surfaces near 10 $\micron$. Using the disc model developed by \citetalias{li2016} as a benchmark model, we have investigated how the mid-IR polarization varies as a function of the disc, dust, and B-field properties. Especially, by `turning off' dichroic emission and absorption in the radiative transfer modeling, we can evaluate the contribution from scattering only. In the context of the present study, we are particularly interested in cases where a protoplanetary disc is unresolved due to the limited sensitivity of an observation. In other words, in such an observation, only the highly compact and bright central region of the disc is detected, which is the case for most, if not all, of our unresolved observations. While detailed results from this investigation is presented elsewhere \citep{zhang2017aas}, key conclusions regarding the scattered polarization are summarized here to aid discussion in the next section.

\begin{figure}
\includegraphics[width=\columnwidth]{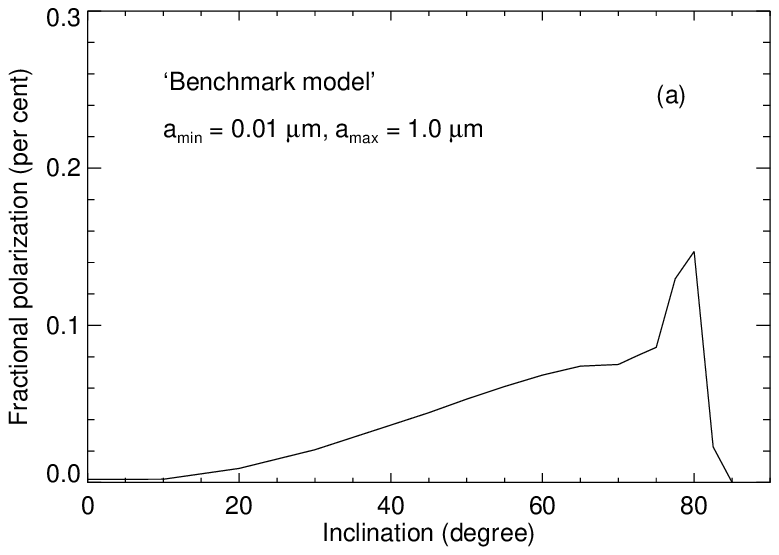}
\includegraphics[width=\columnwidth]{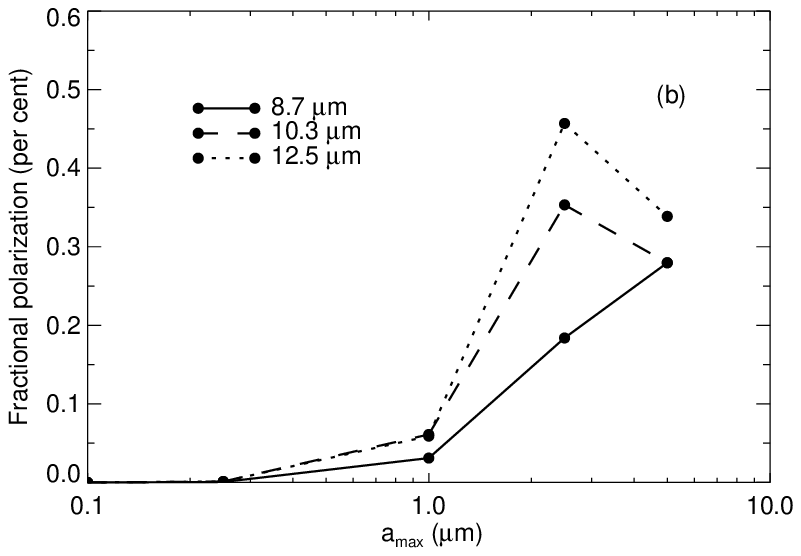} 
\includegraphics[width=\columnwidth]{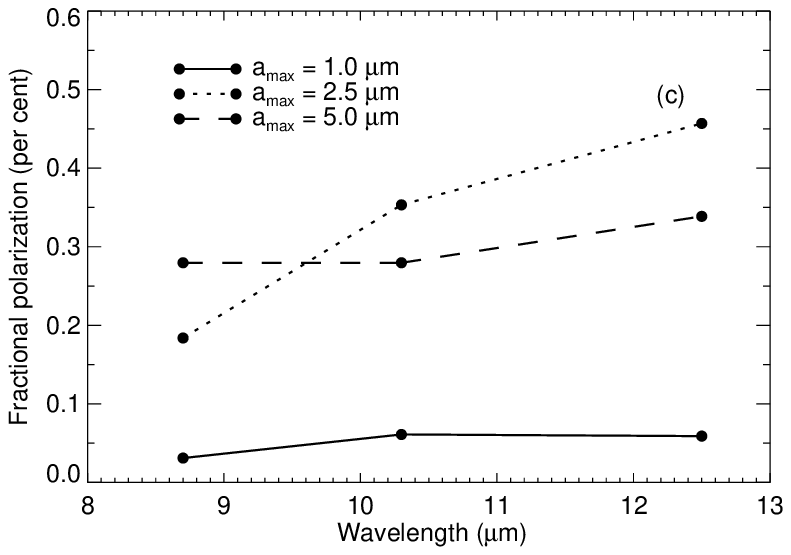} 
\caption{Mid-IR polarization of an unresolved protoplanetary disc. Only polarization produced by scattering is included. Raw model images have been convolved with a PSF kernel to simulate real observations, and the polarization is measured using a circular aperture of 0\farcs35 in radius, corresponding to $\sim$50 au at 144 pc, the distance to AB Aur. $a$: 10.3-$\micron$ polarization measured at different disc inclinations, assuming the benchmark disc model developed for AB Aur \citepalias{li2016}. $b$: Polarization of moderately inclined ($i=45$\degr) discs with different values of $a_{\rm max}$, the maximum grain size. $c$: Polarization measured at different wavelengths for a 45\degr\, inclined disc. 
}
\label{fig:sca_p}
\end{figure}

In the benchmark model, the disc geometry is that used for AB Aur (Table 1 in \citetalias{li2016}). The disc contains only spherical dust particles made of astronomical silicates \citep{draine1984} with a power law dust size distribution. To simulate an unresolved observation, we integrated fluxes from the central ($r<50$ au) part of the model disc and found that the net polarization due to scattering is less than 0.02 per cent in this case. Figure~\ref{fig:sca_p}a shows how the polarization due solely to scattering varies as a function of the disc inclination ($i$). In the benchmark model, the polarization is essentially zero for $i=0\degr$ (i.e., face-on). This is not surprising because a symmetric, face-on disc ensures that polarized signals from different parts of the disc cancel each other out. As the inclination increases, the polarization only increases slightly and remains below 0.1 per cent for $i<70\degr$. Around $i=70\degr$, at which inclination the line of sight to the disc centre approaches the flared disc surface, we observe a sudden rise in the polarization. At $i>80\degr$, most disc surface is blocked by the flared disc itself and the polarization also decreases. This plot suggests that the scattered polarization in the benchmark model is too low to account for the observed mid-IR polarization from a HAeBe disc.

It is known that polarization due to scattering depends strongly on the grain size, and more specifically, the maximum size ($a_{\rm max}$) in the grain size distribution \citep{kataoka2015}. In the benchmark model, we assumed a power law size distribution between $a_{\rm min}$ and $a_{\rm max}$. Figure~\ref{fig:sca_p}b shows that the mid-IR polarization from an unresolved disc increases by nearly an order of magnitude when $a_{\rm max}$ increases from 1 $\micron$ to 2.5 $\micron$ (with $a_{\rm min}$ fixed at 0.01 $\micron$). Hence, scattering by grains of size $\sim$2--5 $\micron$ may be sufficient to explain the level of mid-IR polarization observed in our sample. Note, however, that this scenario requires micron-sized or larger grains to be present at disc surfaces (recall that HAeBe discs are optically thick in the 10-$\micron$ band), whereas one would expect them to have settled to the disc mid-plane \citep[e.g.,][]{laibe2014}. 

Our simulations show that if the mid-IR polarization from a protoplanetary disc is due to scattering, the polarization will be aligned with the minor (spin) axis of the disc regardless of the wavelength. Meanwhile, the degree of polarization will show a wavelength dependence across the 10-$\micron$ band (Fig.~\ref{fig:sca_p}c) different from that expected for polarized emission or absorption. Thus, the wavelength dependence can be used to distinguish between polarization mechanisms, too.

\subsection{Individual sources}\label{sec:individual_source}
\textbf{MWC 1080A} is a Herbig Be star at the centre of a very young cluster \citep{hillenbrand1992,wang2008,li2014}. The disc P.A. (135\degr) is inferred from CO observations, which reveal a bipolar cavity created by the disc outflow \citep{wang2008}. Using near-IR interferometry, however, \citet{eisner2003} yielded a different disc P.A. ($70\degr\pm10\degr$). 

\begin{figure}
\includegraphics[width=\columnwidth]{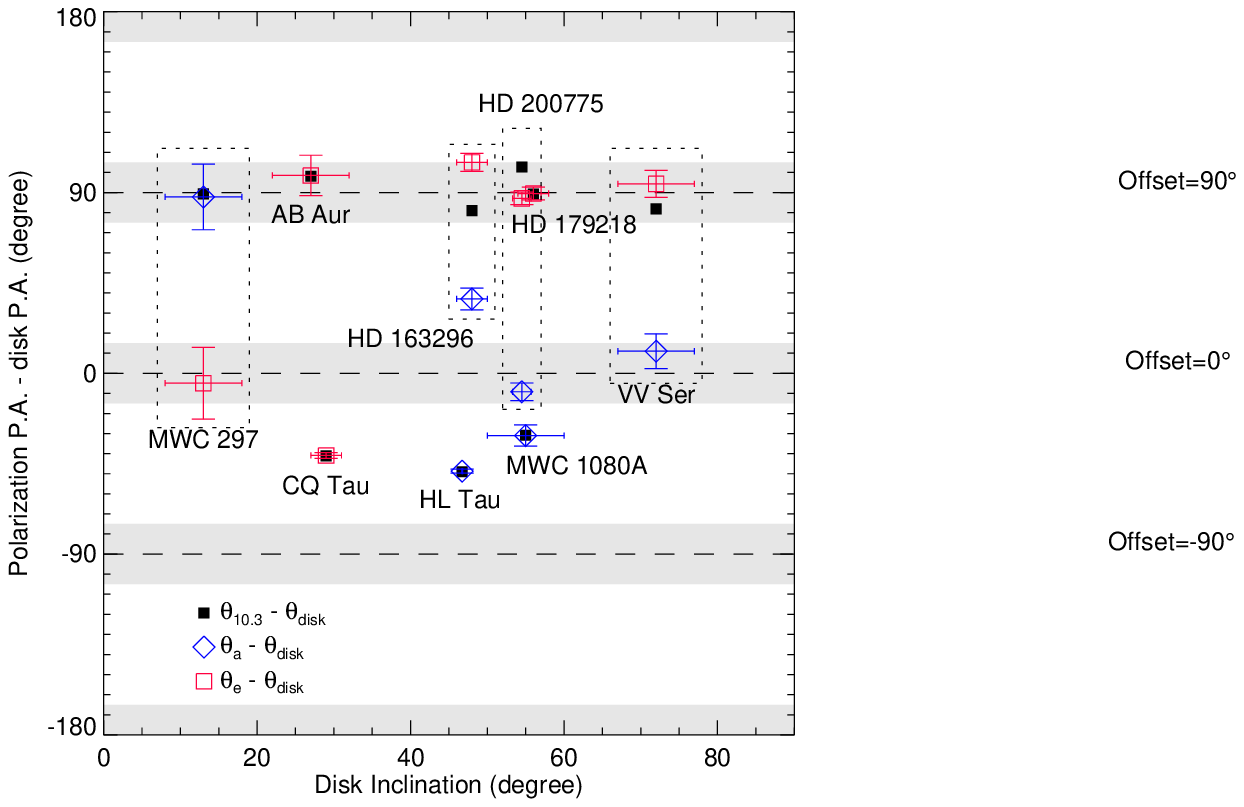} 
\caption{Angle between the disc P.A. and the polarization P.A. The disc P.A. ($\theta_{\rm disc}$) is defined by the major axis of the disc. The polarization P.A. is measured at 10.3 $\micron$ ($\theta_{\rm 10.3}$) or given by the polarimetric fit ($\theta_{\rm a}$ and $\theta_{\rm e}$) as shown in Table~\ref{tab:separation}. If a measurement is found inside gray areas, that means the polarization P.A. is aligned with the major or the spin axis of the disc (within $\pm$15\degr). See the online publication for a colour version of this figure.}
\label{fig:theta_vs_inclination}
\end{figure}

The (error-weighted) mean P.A. of polarization across the 10-$\micron$ band is $107\degr\pm1\degr$. Therefore, neither 135\degr\, nor 70\degr\, seems consistent with polarization due to scattering (Fig.~\ref{fig:theta_vs_inclination}). The polarimetric fit favors the solution of a single absorptive component, but the residual is quite large (Fig~\ref{fig:separation}). However, note that the largest deviation from the model occurs at 12.5 $\micron$, where we only have a marginal (4$\sigma$) detection. For MWC 1080A, we conclude that the polarization is due to absorption arising from the remnant of an envelope or the natal cloud, with the B-field therein likely being neither parallel nor perpendicular to the plane of the disc, as suggested by the offset between the polarization P.A. and the disc P.A.

Similar to MWC 1080A, \textbf{MWC 297}, is among a handful of early B-type stars with compact dust continuum emission from circumstellar discs detected \citep{manoj2007}. Both millimeter interferometry and SED modeling suggest that the disc is close to face-on \citep{manoj2007,alonso-albi2009}. There is also evidence for an envelope surrounding the disc \citep{alonso-albi2009}.

The mid-IR polarization profile of MWC 297 peaks strongly at 10.3 $\micron$ and is fit very well by a two-component model. The polarization is mainly absorptive but also appears to contain some contribution from emission (Fig.~\ref{fig:separation}). Although the polarization P.A. ($80\degr\pm5\degr$) is nearly perpendicular to the disc P.A., this is not necessarily strong evidence for scattering. First, the strongly peaked polarization profile is hard to explain by scattering. Second, the disc P.A. may not be well determined given the very low disc inclination of MWC 297.

The emissive components of the polarization suggests that the projected B-field is aligned with the spin axis of the disc (Fig.~\ref{fig:theta_vs_inclination}). For high-inclination discs, this alignment is often considered a signature of a poloidal field. However, for a nearly-face-on disc like MWC 297, the projection effect is so strong that other field configurations (e.g., a tilted toroidal field) may also produce polarization aligned with the disc axis.

\textbf{HD 200775} is an HBe star associated with the dark cloud L1174 and the illuminating source of the reflection nebula NGC 7023. It is one of the very few HAeBe discs for which flared surfaces have been directly imaged \citep{okamoto2009}. 

HD 200775 has a very peculiar mid-IR polarization profile. The polarization curve is almost flat between 8.7 and 10.3 $\micron$ and rising rapidly towards 12.5 $\micron$. This trend cannot be well reconciled with the polarimetric fit (Fig.~\ref{fig:separation}) but resembles to some degree the curve of scattered polarization assuming $a_{\max}=2.5$ $\micron$ (Fig.~\ref{fig:sca_p}c). The polarization P.A. averaged among three filters is $109\degr\pm1\degr$, approximately perpendicular to the major axis of the disc (P.A. = $6\fdg9\pm1\fdg5$). Hence, we conclude that the mid-IR polarization of HD 200775 is likely due to scattering, although the dichroism scenario cannot be totally ruled out. 

\textbf{VV Ser} is a well-known UX Orionis-type star \citep{rostopchina2001}, which shows large and irregular photometric variability that can be attributed to variable extinction by a rotating, nearly-edge-on disc \citep{dullemond2003}. Wide-field mid-IR images show extended nebulous emission around VV Ser with a wedge-shaped dark lane, which is consistent with a shadow cast by the inner regions of an edge-on disc \citep{pontoppidan2007_vvser2}. Although VV Ser's disc has not been spatially resolved, it must therefore be highly inclined. The orientation of the wedge-shaped dark lane also indicates the disc P.A. \citep{pontoppidan2007_vvser2}.

The polarization profile of VV Ser is similar to that of HD 200775, showing a rise between 10.3 and 12.5 $\micron$, but the profile can still be fit reasonably well by a two-component model (Fig.~\ref{fig:separation}). The P.A. of the emissive component is found to be perpendicular (whereas the absorptive component is parallel) to the disc major axis (Fig.~\ref{fig:theta_vs_inclination}). Considering the high inclination of the disc, a toroidal field is the most probably configuration inferred from our observations. However, the mid-IR polarization of VV Ser is also consistent with scattering: The polarization P.A. is nearly orthogonal to the disc P.A. at all three wavelengths. Perhaps the best way to distinguish between the two scenarios is high-angular resolution and high sensitivity imaging polarimetry using, e.g., ALMA. In the case of pure scattering, our model predicts that the polarization increases rapidly at larger disc radii, reaching $\sim$20 per cent at 50 au, while the net polarization from the entire disc remains below 1 per cent due to self-cancellation. In contrast, the polarization is more uniform if it arises from thermal emission/absorption of aligned grains (assuming uniform B-field and dust properties in the disc). It is also possible that two polarization mechanisms dominate in different regions of the disc, as what was found for AB Aur \citepalias{li2016}. 

\textbf{AB Aur} What would we know about AB Aur if we had not resolved it? To simulate an unresolved observation, we used an oversized aperture to measure the polarization from the entire observed disc \citepalias{li2016}. Our analysis shows that, in this situation, we would likely face the same ambiguity found for VV Ser: On the one hand, the polarization profile can be fit reasonably well by polarized emission (Fig.~\ref{fig:separation}) without invoking scattering. On the other hand, because the polarization P.A. is approximately 90\degr\, offset from the disc P.A., dust scattering would be an equally attractive scenario to explain the data. In \citetalias{li2016}, we resolved this ambiguity by resolving the disc and applying more constraints on the model: The maximum grain size, upon which the degree of scattered polarization strongly depends, was constrained by the polarization observed from the outer part of the disc. Assuming the same dust population, scattering would be too low to account for the polarization from the inner part of the disc, where polarized emission from B-field aligned grains has to be invoked.

\textbf{HD 179218} is surrounded by a flared disc, as suggested by its strong PAH emission and the mm-to-IR SED \citep{acke2004iso}. Although we do not resolve the disc, it is reported to be barely resolvable at 12 $\micron$ with an 8-m telescope \citep{marinas2011}. Using VLTI/MIDI, \citet{fedele2008} obtained the $N$-band visibility of HD 179218. The observation was best fit by a two-ring disc model with an inner ring and an outer ring, of which the inclination ($56\degr\pm2\degr$) is consistent with that inferred from mid-IR nulling interferometry \citep{liu2007}.

HD 179218 is another example for which both scattering and dichroic emission can explain the data reasonably well. The mid-IR light is polarized almost exactly along the minor axis of the disc at all three wavelengths, whereas the degree of polarization increases from 0.41 per cent at 8.7 $\micron$ to about 0.8 per cent at 10.3 and 12.5 $\micron$. These trends are consistent with the scenario of scattering. On the other hand, the polarimetric fit with a single emissive component succeeds in reproducing the observations, too (Fig.~\ref{fig:separation}), implying that the disc of HD 179218 is threaded by a toroidal field.

\textbf{HD 163296} is an isolated HAe star hosting a well-studied protoplanetary disc. The disc PA and inclination have been measured and confirmed repeatedly by observations at multiple wavelengths \citep[e.g.,][]{grady2000,isella2007,tannirkulam2008,qi2011}. We find that it is difficult to reproduce the mid-IR polarization of HD 163296 using emissive and/or absorptive polarization, as shown by the large residual of the polarimetric fit. The polarization profile, which clearly peaks at 10.3 $\micron$, is not consistent with the scattering scenario either, although all of the polarization PAs are approximately along the minor axis of the disc. For this source, we cannot reach a firm conclusion on the origin of its mid-IR polarization. Dust scattering, emission, and absorption may all contribute to the total polarization to some degree.

\textbf{CQ Tau} The disc P.A. and inclination of CQ Tau are derived from CO line observations \citep{chapillon2008}. These measurements are also supported by direct imaging at 20.5 $\micron$ \citep{doucet2006} and interferometric observations and SED modeling \citep{banzatti2011}. The polarization P.A. deviates significantly from the spin axis of disc, suggesting that the process of scattering cannot be the dominant source of polarization for CQ Tau. The polarimetric fit assuming only the emissive component results in a good agreement with the mid-IR data. Because the polarization (and thus the orientation of $B_{\rm pos}$) is not aligned with either the spin or the major axis of the disc, we conclude that the B-field in the CQ Tau disc is not dominated by a toroidal or poloidal component.

\textbf{HL Tau} The 2014 ALMA Long Baseline Campaign revealed astonishing details of the HL disc with unprecedented angular resolution \citep{alma2015}. Compared with the disc P.A. obtained with ALMA, we found that the mid-IR polarization of HL Tau is neither parallel nor perpendicular to the spin axis of the disc (Fig.~\ref{fig:theta_vs_inclination}), thus ruling out scattering as the primary source of polarization. Meanwhile, we found that absorptive polarization fit our observations very well (Fig.~\ref{fig:separation}). 

HL Tau is known to have an unusually massive envelope \citep{beckwith1990}, which is likely to be the place where the mid-IR absorptive polarization arises. Near-IR polarization toward HL Tau is found to be 3.3 and 3.7 per cent in $H$ and $K$ bands, respectively, and has been attributed to dichroic absorption by aligned grains in the envelope \citep{lucas2004}. The polarization P.A. measured in $H$ and $K$ bands (70--80\degr) is also in rough agreement with our measurements in the mid-IR. 

Should the observed near- and mid-IR polarization both arise from the envelope, then these observations imply that the global B-field threading the envelope must be misaligned with the spin axis of the disc. Moreover, the small change in P.A. ($\sim$10\degr) between the near- and mid-IR polarization reflects perhaps a small twist of the B-field along the line of sight, as near- and mid-IR polarization may be produced by two dust populations with different sizes and not well mixed with each other in space.

\citet{stephens2014} spatially resolved the HL Tau disc in polarized light at 1.3 mm, and the polarization vectors were found to be coincident with the spin axis of the disc. Assuming polarization due to emission by aligned dust, \citet{stephens2014} modeled the observation and concluded that the B-field in the HL disc must have a complex structure rather than a simple toroidal or poloidal configuration. It is clear that there is a change ($\sim$50\degr) in the polarization P.A. between the millimeter and IR observations, for which reason we speculate that the B-field configuration must not be the same for the disc and the envelope. However, this is not a firm conclusion. The scenario of polarized emission proposed by \citet{stephens2014} has been challenged by recent studies, which showed that self-scattering occurring in an inclined, optically-thin disc is able to the explain the millimeter polarization from HL Tau \citep{kataoka2015,kataoka2016hltau,yang2016}.

\defcitealias{cho2007}{CL07}

\section{Discussion}\label{sec:discussion}
\begin{figure}
\includegraphics[width=\columnwidth]{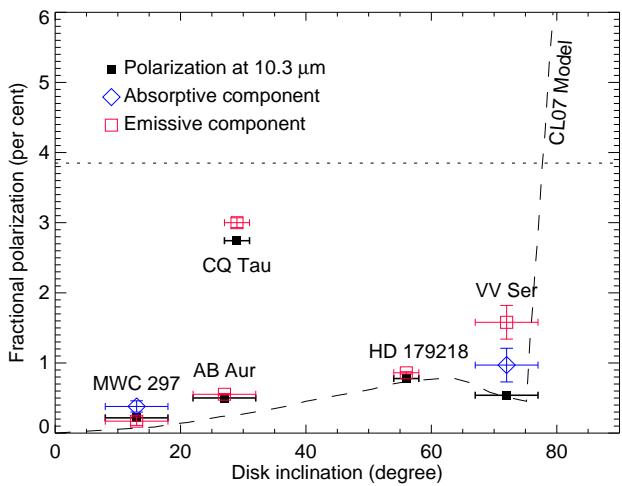} 
\caption{Mid-IR polarization of the sample. The dash line is polarized emission at 10 $\micron$ predicted by \citet{cho2007} for a flared disc, assuming a toroidal B-field and RAT alignment. The horizontal line is the mean 10-$\micron$ polarization of YSOs in \citet{smith2000}. See the online publication for a colour version of this figure.}
\label{fig:p_vs_i}
\end{figure}

Polarization arising from aligned grains in a disc is a highly complex process. In order to produce quantitative results, theoretical and numerical studies usually assume a highly simplified dust model (i.e., one dust size and shape) and/or fixed dust alignment efficiency everywhere in the disc \citep[e.g.,][]{aitken2002}. More sophisticated theories such as RAT are able to predict whether grains of a given size and shape will be aligned under certain conditions (field strength, gas density, etc.), thus yielding more realistic estimate for disc polarization. For example, \citet[][hereafter \citetalias{cho2007}]{cho2007} computed polarized IR and millimeter emission from a flared protoplanetary disc, assuming a toroidal B-field and dust alignment by RAT. In Fig.~\ref{fig:p_vs_i}, we compare \citetalias{cho2007}'s prediction to our observations (we only plotted five objects for which the presence of polarized emission has been suggested by the observations) and we see that the polarization from all objects except CQ Tau is consistent with \citetalias{cho2007}'s model. 

This agreement is in contrast to most other follow-up observations of \citetalias{cho2007} carried out in the millimeter, yielding either non-detection or polarization much lower than \citetalias{cho2007}'s predictions \citep[e.g.,][]{hughes2008,krejny2009,rao2014}. To explain the discrepancy, a few mechanisms were discussed, e.g., an over-estimated dust alignment efficiency or inaccuracies in \citetalias{cho2007}'s disc model \citep{hughes2009, krejny2009}. Recently, \citet{tazaki2017} revisited the issue of grain alignment by RAT in protoplanetary discs, and they pointed out that large grains ($\sim$a few hundred microns) at the disc mid-plane are difficult to align due to the efficient gaseous damping as well as fast radiative precession. Meanwhile, small, sub-micron-sized grains can be aligned with the B-field at the surface layer of the disc where the gas density is lower. Because those small grains at the disc surface account for most of the disc mid-IR emission, while larger grains in the mid-plane account for most of the mm/sub-mm emission, the findings of \citet{tazaki2017} may explain why \citetalias{cho2007} correctly predicts mid-IR polarization while overestimating mm/sub-mm polarization from discs.

Regarding the apparent inconsistency between the observation of CQ Tau and \citetalias{cho2007}'s predictions, although it is difficult to reach any conclusion without further observation and modeling, we can think of several possible explanations. For example, the unusually high polarization may result from an exceptionally high dust alignment efficiency in the disc of CQ Tau. Or, the disc inclination of CQ Tau may have been considerably underestimated. CQ Tau is known to exhibit UX Ori-like variability \citep{natta1997}, which only occurs when the light of sight is close to the plane of the disc. Despite that the low disc inclination is not well supported by direct or interferometric imaging \citep[e.g.,][]{chapillon2008,banzatti2011}, it is possible that the unresolved inner disc of CQ Tau (which is also the part of the disc where most mid-IR polarization arises) is heavily warped with respect to the outer disc.

There are seven objects in our sample of which the observations are consistent with dichroic emission and/or absorption (Table~\ref{tab:origins}). Note that here we have excluded HD 200775 and HD 163296, for which no solid conclusion regarding the origin of polarization has been drawn. The detection rate (seven out of eleven, or 64 per cent) of dichroic emission/absorption is similar to that found for a much larger sample of Class 0/I YSOs \citep{smith2000}. However, this agreement might be a coincidence, considering the differences between the two samples: objects studied by \citet{smith2000} are in general younger and significantly more massive than ours. Nevertheless, we conclude that, even for relatively evolved (compared to Class 0/I YSOs) objects like HAeBe discs, B-fields and suitable conditions for effective dust alignment are still present at disc surfaces.

\begin{table*}
\caption{Origin of polarization.}
\label{tab:origins}
\begin{tabular}{lcccc} 
\hline
Object & Scattering & Emission & Absorption & B-field\\
\hline
MWC 1080A &  &  & $\surd$ & Complex or tilted \\
MWC 297 &  & $\surd$ & $\surd$ & Poloidal$^a$ \\
HD 200775 & $\surd$ &  & \\
VV Ser & $\surd$ & $\surd$ & $\surd$ & Toroidal \\
HD 179218 & $\surd$ & $\surd$ & & Toroidal \\
AB Aur &  $\surd$ & $\surd$ & & Tilted poloidal \citepalias{li2016} \\
HD 163296 &  & & & \\
CQ Tau &  & $\surd$ & & Complex or tilted \\
HL Tau &  &  & $\surd$ & Complex or tilted \\
\hline
\multicolumn{5}{l}{Possible origin(s) of polarization. See Section \ref{sec:individual_source} for details.}\\
\multicolumn{5}{l}{$^a$ Poorly constrained. See Section \ref{sec:individual_source}.}\\
\end{tabular}
\end{table*}

The mid-IR polarization of HAeBe discs is lower than that found for YSOs (Fig.~\ref{fig:p_vs_i}). This can be explained, to some degree, by the generally higher column density (optical depth) toward most YSOs, and the absorptive polarization increases with increasing extinction. The difference may also be partially due to the change in dust properties and/or the B-field strength. Dust properties and disc environments may evolve in such a way that the dust alignment efficiency decreases with time. For example, because only sub-micron grains contribute to polarized emission in the mid-IR while grains larger than a few microns only emit unpolarized light, grain growth can be an effective process to reduce mid-IR polarization of discs \citep{tazaki2017}.

The alignment (or misalignment) between the large-scale B-field and the disc spin axis may play an important role in the disc evolution. For example, some misalignment between the two could mitigate the so-called `magnetic braking catastrophe' \citep{allen2003,mellon2008,hennebelle2009,joos2012,joos2013,seifried2013,li2013}. In the present study, we do not see a clear trend that the B-field in HAeBe discs tends to align with the spin axis or the plane of the disc: most objects in our sample seem to have a complex rather than poloidal or toroidal B-field configuration. This is in agreement with a few previous studies where B-fields were also found to be uncorrelated with the disc structure \citep{curran2007,rao2009, shinnaga2012, hull2013, zhang2014}. Theoretical studies and numerical simulations have revealed that the B-field structure can be highly complex in circumstellar discs \citep{tomisaka2011,kataoka2012, dudorov2014, seifried2015}. In the presence of turbulence, outflow, and/or disc rotation, the B-field structure can become very disordered in the vicinity or the central region of a protoplanetary discs \citep{seifried2015,tomisaka2011}. Hence, the alignment between core and disc fields, even if it exists at large scales (i.e., a few thousand au), may not persist at smaller disc scales (i.e., a few hundred au or tens of au).

We have compared (projected) orientations of disc B-fields to interstellar B-fields (the sixth column in Table~\ref{tab:angles}). Despite a few `aligned' cases, being HL Tau, AB Aur, and perhaps HD 179218 (for disc B-field orientation derived from the absorptive, emissive, and emissive component, respectively), we do not see a strong correlation between the two. Such a correlation may exist if disc B-fields are `fossil' ones from their parent molecular clouds, of which the B-fields are the frozen-in fields from the ISM of the Milky Way \citep{crutcher2012}. However, under influences from accretion, magnetic diffusion, and different instabilities, such a memory of the parent interstellar field may have already been lost.

\section{Summary}\label{sec:summary}
We detected linear polarization at 8.7, 10.3, and 12.5 $\micron$ from eight Herbig Ae/Be stars and one T-Tauri star. We fit the polarization profile for each object in the 10-$\micron$ band to a combination of polarized emission and absorption. We also considered the role of scattering in the interpretation of the data. While the mid-infrared polarization of most objects is consistent with polarized emission and/or absorption arising from aligned grains, we cannot rule out the scattering scenario for a few objects in our sample. For those objects with mid-IR polarization being consistent with polarized emission and/or absorption, we examined how the inferred magnetic field structure correlates with the viewing geometry (inclination and P.A.) of the disc. We found no preference for a certain configuration (with respect to the spin axis of the disc) of the magnetic field. Instead, various geometries (toroidal, poloidal, or complex) may apply to different objects. The detection rate (seven of eleven, or 64 per cent; excluding objects for which the scattering scenario is a more consistent with the data) of mid-infrared polarization due to emission and/or absorption suggests that magnetic fields and suitable conditions for grain alignment should be common in protoplanetary discs around Herbig Ae/Be stars.

\section*{Acknowledgements}
We are grateful to the GTC staff for their support during the queue observations. E.P. acknowledges the support from the AAS through the Chr\'{e}tien International Research Grant and the FP7 COFUND program-CEA through an enhanced-Eurotalent grant, and the University of Florida for its hosting through a research scholarship. C.M.T. acknowledges support from NSF grants AST-0908624, AST-0903672, and AST-1515331. C.M.W. acknowledges support from an Australian Research Council Future Fellowship, FT100100495. This research is based on observations using CanariCam at the Gran Telescopio Canarias (GTC), a partnership of Spain, Mexico, and the University of Florida, and located at the Spanish Observatorio del Roque de los Muchachos of the Instituto de Astrof\'{i}sica de Canarias, on the island of La Palma.

\bibliographystyle{mnras}
\bibliography{references} 

\bsp	
\label{lastpage}
\end{document}